\documentclass[11pt,a4paper]{emulateapj}
\usepackage{epsfig}
\usepackage{amsmath}
\usepackage{natbib}
\usepackage{color}
\usepackage{wasysym}

\begin{document}

\title{Stellar wind induced soft X-ray emission from close-in exoplanets}
\author{K.G.~Kislyakova$^{1}$, L.~Fossati$^{2}$, C.P.~Johnstone$^{3}$, M.~Holmstr{\"o}m$^{4}$, V.V.~Zaitsev$^{5}$, \& H.~Lammer$^{1}$}
\affil{$^{1}$Space Research Institute, Austrian Academy of Sciences, Graz, Austria; kristina.kislyakova@oeaw.ac.at\\
$^{2}$Argelander-Institut f{\"u}r Astronomie der Universit{\"a}t Bonn, Bonn, Germany\\
$^{3}$University of Vienna, Department of Astrophysics, Vienna, Austria\\
$^{4}$Swedish Institute of Space Physics, Kiruna, Sweden \\
$^{5}$Institute of Applied Physics, Russian Academy of Sciences, Nizhny Novgorod, Russia
}
\date{\today}

\begin{abstract}
In this paper, we estimate the X-ray emission from close-in exoplanets. We show that the Solar/Stellar Wind Charge Exchange Mechanism (SWCX) which produces soft X-ray emission is very effective for hot Jupiters. In this mechanism, X-ray photons are emitted as a result of the charge exchange between heavy ions in the solar wind and the atmospheric neutral particles. In the Solar System, comets produce X-rays mostly through the SWCX mechanism, but it has also been shown to operate in the heliosphere, in the terrestrial magnetosheath, and on Mars, Venus and Moon. Since the number of emitted photons is proportional to the solar wind mass flux, this mechanism is not very effective for the Solar system giants. Here we present a simple estimate of the X-ray emission intensity that can be produced by close-in extrasolar giant planets due to charge exchange with the heavy ions of the stellar wind. Using the example of HD~209458b, we show that this mechanism alone can be responsible for an X-ray emission of $\approx 10^{22}$~erg~s$^{-1}$, which is $10^6$ times stronger than the emission from the Jovian aurora. We discuss also the possibility to observe the predicted soft X-ray flux of hot Jupiters and show that despite high emission intensities they are unobservable with current facilities.

\end{abstract}

%Section heading
\section{Introduction}

%\subsection{X-ray emission from the bodies of the Solar system: mechanisms and observations}

X-ray emission has been observed for many of the Solar system objects, e.g. for Mars \citep{Holmstroem2001,Gunell2004,Dennerl2002}, Venus \citep{Bhardwaj2007,Dennerl2002b}, Earth and the Moon \citep{Collier2014,Bhardwaj2007}, Jupiter and the Galilean satellites \citep{Cravens2003,Bhardwaj2007}, Saturn \citep{Branduardi-Raymont2010}, comets \citep{Cravens2002,Lisse2004}, and in the heliosphere \citep{Cravens2001}.

For the Solar system planets, X-rays are known to be generated via different mechanisms. The main mechanisms are:

-- continuum Bremsstrahlung emission due to collisions with electrons (produces mostly hard X-rays);

-- excitation of neutral species and ions due to collisions, e.g., with electrons (charged particle impact), followed by line emission;

-- stellar X-ray photon scattering from neutrals in planetary atmospheres (elastic scattering and K-shell fluorescent scattering, requires a significant column density);

-- charge exchange between the solar wind ions with neutrals (SWCX), followed by X-ray emission;

-- X-ray production from the charge exchange of energetic (energies of about a MeV/amu) heavy ions of planetary magnetospheric origin with neutrals or by direct excitation of ions in collisions with neutrals (this is known to be effective on Jupiter, e.g. \citealp{Cravens2003,Bhardwaj2007}).

The cross sections at solar wind energies for charge exchange with the solar wind heavy ions are several magnitudes larger than the cross sections for the excitation of the neutral species by electrons \citep{Bhardwaj2007}, which makes the SWCX mechanism more effective.

In the present article, we discuss the SWCX mechanism and X-ray scattering as applied to close-in giant exoplanets, in particular to HD~209458b. We discuss the observability of the X-ray emission from HD 209458b, X-ray emission from other giant planets, and the influence of the host star age. Other X-ray production mechanisms are beyond the scope of the present article, but will be the goal of a future study.

%The last mechanism is known to be effective on Jupiter \citep{Cravens2003,Bhardwaj2007}, but requires a comprehensive study and is beyond the scope of the present article.

%\begin{enumerate}
%	\item continuum bremsstrahlung emission produced by collisions with electrons;
%	\item excitation of neutral species and ions due to collisions, e.g., with electros (charged particle impact), followed by line emission;
%	\item stellar X-ray photon scattering from neutrals in planetary atmospheres (elastic scattering and K-shell fluorescent scattering, requires a significant column density);
%	\item charge exchange between the solar wind ions with neutrals, followed by X-ray emission;
%	\item X-ray production from the charge exchange of energetic (energies of about a MeV/amu) heavy ions of planetary magnetospheric origin with neutrals or by direct excitation of ions in collisions with neutrals.
%\end{enumerate}

\section{Solar wind charge exchange mechanism}

%The description of the mechanism itself. References to Jupiter, Saturn, comets.

In the SWCX mechanism, an electron is transferred from a neutral atom or molecule to a highly charged heavy ion of the solar wind. This mechanism is known to produce soft X-rays in cometary comas \citep{Cravens2002}. In the case of a magnetized planet, these ions can enter the neutral atmosphere following the open field lines near the polar cusp. 

It is known from experimental and theoretical studies that solar wind heavy ions can undergo charge exchange reactions when they are within approximately 1 nm of a neutral atomic species (e.g., \citealp{Lisse2004,Cravens2002,Bhardwaj2007} and references therein):
\begin{equation}
 {\rm A^{q+} + B \rightarrow A^{(q-1)+*} + B^+},
 \label{e_cx}
\end{equation}
where A is a charged heavy ion in the solar wind (the projectile), q is the projectile charge and B is a neutral component (target). The product ion $\rm A^{(q-1)+*}$ is still highly charged and is almost always left in an excited state (marked by an ``*"). Then, the excited ion emits one or several X-ray photons in the following reaction:
\begin{equation}
 {\rm A^{(q-1)+*} \rightarrow A^{(q-1)+} +} h \nu.
 \label{e_deex}
\end{equation}
Although the de-excitation usually represents a number of cascading processes through intermediate states, if q is high then an X-ray photon (at least one, though usually several) is emitted \citep{Cravens2002}.

The composition of the solar wind by volume is 0.92 hydrogen, 0.08 helium, and $\approx 10^{-3}$ heavier elements. Since the solar wind quickly becomes collisionless as it expands, the charge states that the heavy ions have in the hot solar corona are frozen-in, and therefore the heavy elements are usually highly charged. In the solar wind, the most common heavy ions are C$^{4+}$, C$^{5+}$, N$^{6+}$, O$^{6+}$, O$^{7+}$, O$^{8+}$, Ne$^{8+}$, Si$^{9+}$, Fe$^{12+}$. %One of the possible charge exchange reactions is

%\begin{equation}
% {\rm O^{7+} + H \rightarrow O^{6+*} + H^+},
% \label{e_cx2}
%\end{equation}
%where O$^{6+*}$ is an excited state of the O$^{6+}$ ion and emits at least one X-ray photon. 
The cross sections for such charge transfer collisions are very high at solar wind energies, exceeding 10$^{15}$~cm$^2$ (e.g., \citealp{Greenwood2001}). The types of the species that undergo charge exchange define the energy of the emitted X-ray photons, which is usually in the range of 0.3--0.5~keV.

\section{SWCX mechanism on hot Jupiters: the case study of HD~209458b}

%Hot Jupiters stuff. Discussion about aurora, magnetosphere, stellar wind and so on. Flux, etc, etc.
%Estimate on an example of HD~209458b. Cite \cite{Kislyakova2014} about stellar wind parameters and magnetic moment of the planet $\rightarrow$ aurora size.
%--------------------------------------------------
\begin{table}
  \caption{Planetary and stellar wind parameters, HD~209458b}
  \begin{center}
    \leavevmode
    \begin{tabular}{ll} \hline \hline              
  Name										& Value\footnote{adopted from \cite{Kislyakova2014}}     \\ \hline 
  Planetary mass, $M_p$						& $\approx 0.71 M_{\jupiter}$     \\
  Planetary radius, $R_p$					& $\approx 1.38 R_{\jupiter}$     \\
  Semi-major axis							& $\approx 0.047$~AU         	 \\
  Stellar wind velocity, $v_{\rm sw}$		& $4 \times 10^7$~cm~s$^{-1}$    \\
  Stellar wind density, $n_{\rm sw}$		& $5 \times 10^3$~cm$^{-3}$       \\
  Fraction of heavy ions\footnote{solar wind value assumed}, $f$				& $10^{-3}$		 \\
  Planetary magnetic moment, $\mathcal{M}_{p}$ & $\approx 0.1\mathcal{M}_{\jupiter}$		\\ 
  Magnetospheric stand-off distance, $R_s$	& $2.9 R_p$        				\\ \hline
    \end{tabular}
  \end{center}
   \label{t1}
\end{table}
%--------------------------------------------------

In this section, we discuss the soft X-ray emission which can be produced by the SWCX mechanism on close-in exoplanets. As an example, we consider HD~209458b, which is a well-studied close-in gas giant orbiting a 4$\pm$2~Gyr old G-type star. Planetary and stellar wind parameters are summarized in Table~\ref{t1}. In our further estimations, we rely on results of \cite{Kislyakova2014}, who investigated the magnetosphere and stellar wind parameters in the vicinity of HD~209458b by means of modelling. Their result support a magnetic moment of HD~209458b of approximately 10\% of the Jupiter's and a stellar wind with a velocity of $4 \times 10^7$~cm~s$^{-1}$ at the time of observation.

For a very simple estimate of the X-ray intensity, $I$, emitted in the region of the atmosphere exposed to heavy ion precipitation one can use the following expression \citep{Cravens2003}:
\begin{equation}
	4 \pi I \approx 2 n_{\rm sw} v_{\rm sw} f N,
     \label{e_int}
\end{equation}
where N is a factor of 2 or 3 and represents the number of photons emitted per ion (below we assume $N=3$). The additional factor of 2 on the right hand side is a flank magnetosheath enhancement factor \citep{Cravens2003}.

%Here 1~Rayleigh (1~R) is a unit of $4 \pi I$ which equals $10^6$~cm$^{-2}$s$^{-1}$, where the intensity, $I$, has units of cm$^{-2}$s$^{-1}$sr$^{-1}$.

For Jupiter, Equation \ref{e_int} yields $4 \pi I \approx 10^5$~cm$^{-2}$ s$^{-1}$ while values of $2\times10^6 - 2\times10^7$~cm$^{-2}$ s$^{-1}$ are necessary to explain the observed auroral soft X-ray emission, which means that SWCX is not the main mechanism that produces soft X-ray emission for Jupiter.

For HD~209458b, we assume the stellar wind parameters estimated by \cite{Kislyakova2014}, $n_{\rm sw} = 5 \times 10^3$~cm$^{-3}$, $v_{\rm sw} = 4 \times 10^7$~cm~s$^{-1}$ and $f \approx 10^{-3}$ which gives  $4 \pi I \approx 1.2 \times 10^9$~cm$^{-2}$ s$^{-1}$ or 60--600 times the observed Jovian values.

Given its proximity to its host star, it is unclear whether HD 209458b is located in the sub-Alfv$\acute{\rm e}$nic or super-Alfv$\acute{\rm e}$nic region of the wind. The exact regime depends on the magnetic moment of the host star and stellar wind parameters. Although the results of \cite{Kislyakova2014} support that HD~209458b is rather in the super-Alfv$\acute{\rm e}$nic regime and thus outside the stellar plasma corotation region, we make an estimate also for the corotation case. HD~209458 has been observed to have a rotational velocity of 4.4~km/s which corresponds to a rotational period of $\approx$11.5 days \citep{Mazeh2000}. This gives the corotational velocity of plasma at 0.047~AU of $v_{\rm cor} \approx 2.7 \times 10^{6}$~cm~s$^{-1}$. Taking into account also the Keplerian orbital speed $v_{\rm orb}\approx1.4\times10^7$~cm~s$^{-1}$, this corresponds to the plasma flow velocity in the vicinity of HD~209458b of $v_{\rm flow} \approx 1.2 \times 10^7$~cm~s$^{-1}$. Substituting it into the equation \ref{e_int} instead of $v_{\rm sw}$, one obtains an estimate of $4 \pi I \approx 3.6 \times 10^8$~cm$^{-2}$ s$^{-1}$, which is still 18--180 times the observed Jovian value.

To estimate the aurora size of HD~209458b we follow the approach of \cite{Vidotto2011}. The fractional area of the planetary surface that has open magnetic field lines is $(1-\cos \alpha_0)$ for both the north and south auroral caps, where
\begin{equation}
	\alpha_0 = \arcsin[(R_p/R_s)^{1/2}],
     \label{e_aurora}
\end{equation}
$R_p$ is the radius of a planet, and $R_s$ is the magnetosphere stand off distance at the substellar point. Assuming $R_s = 2.9 R_p$ estimated by \cite{Kislyakova2014}, we obtain $\alpha_0 \approx 0.63$ and $1-\cos \alpha_0 \approx 0.19$. This gives the size of the aurora of $A \approx 2.17 \times 10^{20}$~cm$^2$, or $\approx 217$ times the Jovian aurora $A_{\jupiter} \approx 10^{18}$~cm$^2$ \citep{Cravens2003}.

Now we can estimate the power of the soft X-ray emission from HD~209458b in both the corotation and non-corotation regimes. For simplicity, we assume the energy of each emitted X-ray photon is 0.3 keV \citep{Cravens2003}. Using the solar value of $f = 10^{-3}$, we estimate the total X-ray power of HD~209458b to be $\approx 1.3 \times 10^{20}$~erg~s$^{-1}$ in the non-corotation regime (point C on Fig.\ref{f1}) and $\approx 2.3 \times 10^{19}$~erg~s$^{-1}$ in the corotational regime (Fig.\ref{f1}, point D).

Note that these values can still present a lower limit. If charge exchange occurs not only in the auroral regions of HD~209458b, but in the whole hemisphere with the radius of $R_s$, the values should be multiplied by a factor of $\approx 88$, which is the ratio of the interaction area sizes. This gives $\approx 1.1 \times 10^{22}$~erg~s$^{-1}$ in the non-corotation regime (point A in Fig.\ref{f1}) and $\approx 2.0 \times 10^{21}$~erg~s$^{-1}$ in the corotation regime (Fig.\ref{f1}, point B). Since the atmospheres of hot Jupiters in general and HD~209458b in particular are highly inflated and are believed to extend beyond the magnetosphere \citep{Kislyakova2014}, this is a more realistic case than interaction only in the aurora region (Jupiter type). The X-ray production can be even larger if the region outside the magnetosphere (the volume between the magnetopause and the bow shock) is included \citep{Robertson2003}.

\subsection{Contribution of stellar X-ray photon scattering}

Although only a small fraction of the incident stellar X-rays are reflected by the planetary atmosphere, in the Solar system this mechanism is known to contribute to the total soft X-ray luminosity of planets and dominates, for example, the X-rays from Venus \citep{Dennerl2002b}. \cite{Cravens2006} showed that the scattering albedo for the outer planets is quite small and equals $10^{-3}$ at 3~nm. We assume this albedo for HD~209458b as a crude estimate.

The total X-ray luminosity of HD~209458 was first observed to be $\log L_{\rm X} \approx 27.02 \pm 0.2$~erg~s$^{-1}$ \citep{Kashyap2008}. However, later it was reported that this result may present a luminosity of a nearby star and a new upper limit of $\log L_{\rm X} \lesssim 26.12$~erg~s$^{-1}$ was reported \citep{Sanz-Forcada2010}. Both values are close to the Solar X-ray luminosity of $\log L_{\rm X \odot} \approx 26-27$~erg~s$^{-1}$.

The value of $\log L_{\rm X} = 26.12$~erg~s$^{-1}$ yields an X-ray flux of $\approx 39$~erg~cm$^{-2}$~s$^{-1}$ in the vicinity of HD~209458b and an X-ray luminosity of reflected soft X-rays from the planet of $\approx 2.3 \times 10^{19}$~erg~s$^{-1}$,  which is comparable only to our lowest estimate for the X-ray flux produced by the SWCX mechanism (point D in Fig\ref{f1}).

%Altogether, the highest value of the soft X-ray luminosity of HD~209458b due to SWCX mechanism and the photon scattering is $\approx 1.1 \times 10^{22}$~erg~s$^{-1}$.

\section{Observability}
In order to see whether the soft X-ray exoplanetary emission described above could be observable with currently available facilities, we have considered the maximum X-ray luminosity estimated for HD~209458b (i.e., $1.1 \times 10^{22}$~erg~s$^{-1}$) and that of its host star ($\log L_{\rm X}=27.02$~erg~s$^{-1}$; \citealp{Kashyap2008}). Note that \cite{Sanz-Forcada2010} reported only an upper limit of $\log L_{\rm X}<26.12$~erg~s$^{-1}$ on the X-ray luminosity of HD~209458 and concluded that \cite{Kashyap2008} might have confused the star with a nearby object. For our purposes we are interested in the best possible scenario and therefore assume the X-ray luminosity given by \cite{Kashyap2008}. A lower stellar X-ray luminosity would make the detection of the planetary X-ray emission harder to detect compared to what described here. By rescaling both luminosities to the distance of HD~209458 ($d=49.6$~pc; \citealp{vanLeeuwen2007}), we obtained that the maximum X-ray flux of HD~209458b would be $3.9 \times 10^{-20}$~erg~cm$^{-2}$~s$^{-1}$, while from the star we would get an X-ray flux of $3.6 \times 10^{-15}$~erg~cm$^{-2}$~s$^{-1}$\footnote{The slight difference in the stellar X-ray flux with that given by \cite{Kashyap2008} is due to the use of a different distance.}, resulting in a difference of about 5 orders of magnitude.

The planetary X-ray emission could be observed using the secondary transit, as commonly done in the infrared, for example. The ratio between the in-transit and out-of-transit fluxes is expected to be of the order of 10$^{-5}$~erg~cm$^{-2}$~s$^{-1}$, which would require a signal-to-noise ratio (S/N) of the measurements of the order of 10$^5$ to be detected. Such a high precision is currently reached in the optical and infrared bands (mostly with space observations), but it is prohibitive at X-ray wavelengths. To highlight this, we used the available count rate simulator\footnote{\tt http://heasarc.gsfc.nasa.gov/cgi-bin/Tools/w3pimms/w3pimms.pl} for the XMM-Newton telescope that, among the facilities currently available, has the largest efficiency in the soft X-rays. Taking into account the 0.1--1.0~keV band, a plasma with a temperature of 10$^6$~K, and the count rate given for the pn detector and the ``thin'' filter we obtained a count rate of $3.9 \times 10^{-3}$~counts~s$^{-1}$ for the X-ray emission of HD~209458. As a result, the S/N obtained exposing for $10^3$~seconds is about 2 and it would require several thousand years of exposure time to reach the S/N required to detect the planetary X-ray emission. The situation might slightly improve if the planetary X-ray emission has a different spectral behaviour compared to that of the star, but the detectability would probably remain unfeasible. Here we have not considered the intrinsic stellar X-ray variability which will hamper the detection of the planetary X-ray emission.

\section{SWCX mechanism on other giant exoplanets}

%-----------------------------------------------------
% Example of figure: force single column format
\begin{figure*}
  \begin{center}
		\includegraphics[width=2.0\columnwidth]{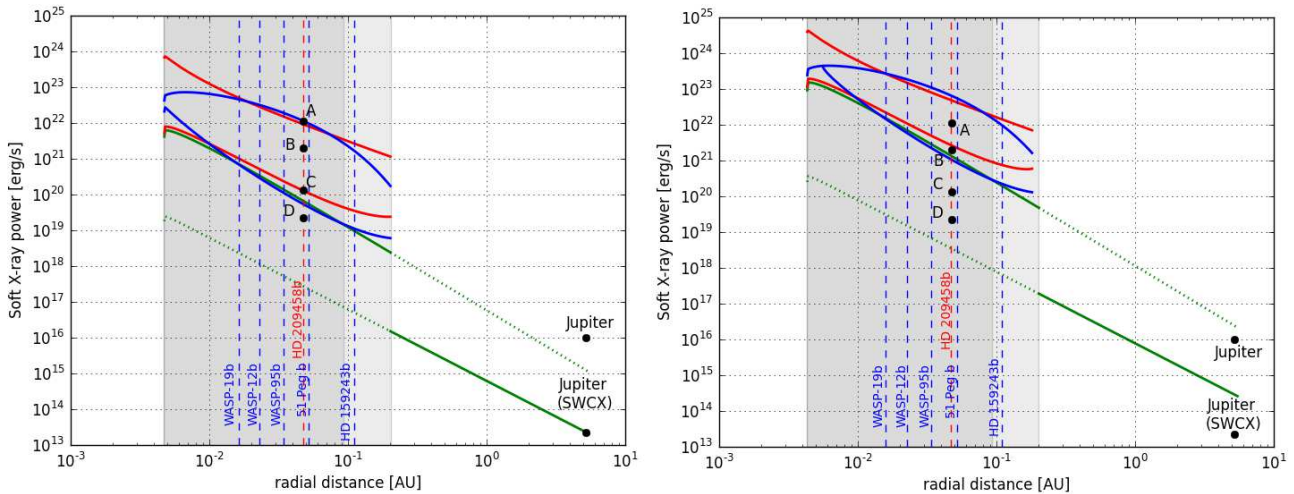}
        \caption{Dependence of the soft X-ray emission produced by the SWCX mechanism on the orbital semi-major axis of a HD~209458b-like giant planet. Left panel: the calculation using the stellar wind parameters $f$, $n_{\rm sw}$, $v_{\rm sw}$ of the current solar wind (4.5 Gyr old G2V dwarf). The bright shaded area shows the locations where planets could potentially be tidally locked and the dark shaded area shows where planets will be tidally locked. The letters A, B, C, and D mark the estimates for HD~209458b. The upper and lower green lines are estimates for emission from auroral region with aurora size calculated according to Eq.~\ref{e_aurora} and for a fixed Jovian aurora of $A_{\jupiter} = 10^{18}$~cm$^2$ respectively. The upper and lower red lines are estimates for a tidally locked planet for the charge exchange in the whole hemisphere restricted by $R_s$ and only auroral regions respectively. The upper and lower blue lines illustrate the estimates for a tidally locked planet in a corotation regime. The vertical dashed blue lines show the semi-major axes of some known hot Jupiters orbiting G dwarf stars. The point labelled ``Jupiter" marks the observed soft X-ray emission from the Jovian aurora, and ``Jupiter (SWCX)" stands for the intensity produced via the SWCX mechanism only. Right panel: the same as the left panel, but assuming the stellar wind parameters of the young stellar analogue EK Dra (100~Myr old G1.5V dwarf). }
     \label{f1}
  \end{center}
\end{figure*}
%-----------------------------------------------------

In this section, we briefly consider the influence of the orbital distance on the soft X-ray emission generated by the SWCX mechanism (the radius of HD~209458b is assumed). For the stellar wind parameters, we consider two cases: for one case, we assume the wind has the properties of the slow component of the current solar wind, and for the other case, we scale the wind properties to match what we might expect for the young solar analogue EK Dra.

We calculate the solar slow wind parameters as a function of distance from the star using a 1D hydrodynamic wind model that is constrained by \emph{in situ} spacecraft measurements of the real solar wind. The model was developed by \cite{Johnstone2015a} and provides a very good description of the real solar wind outside of the solar corona. 

Although little is known about the properties of winds from other stars, it is suspected that more active stars have mass fluxes that are significantly higher than the mass flux of the current solar wind \citep{Wood2005,Holzwarth2007,Suzuki2013}.
To scale the slow solar wind model to EK Dra, we use the scaling relation for mass loss rate derived by \cite{Johnstone2015b} and the parameters for EK Dra given by \cite{Guedel2007}.

We find a mass loss rate, and therefore corresponding values for $n_\text{sw} v_\text{sw}$, for EK Dra that are approximately a factor of 15 higher than in the current solar wind.

Based on the example of the solar wind, we might expect that the abundances of heavy ions are similar in the corona and in the wind. While
it is known that the coronal abundances are correlated with coronal activity for Sun-like stars, the coronal abundances of the most active
stars are only approximately a factor of two different from the solar values \citep{Telleschi2005}. Since this is an insignificant difference compared to all other uncertainties, for simplicity we assume a solar value of $f \approx 10^{-3}$.

Magnetic moments of tidally locked gas giants are believed to be smaller than $\mathcal{M}_{\jupiter}$ because of their slower rotation \citep{Griessmeier2004,Khodachenko2012}. For HD~209458b, this hypothesis was lately also supported by a modelling result based on the Ly$\alpha$ transit observations \citep{Kislyakova2014} which predicted a planetary magnetic moment of $\mathcal{M}_{p} \approx 0.1\mathcal{M}_{\jupiter}$. Although a prediction of magnetic moments of hot Jupiters $\ge \mathcal{M}_{\jupiter}$ also exists \citep{Christensen2009}, in the present study we assume a moment value in the range of $\approx 0.05 - 0.5\mathcal{M}_{\jupiter}$ (see Fig.~2 in \citealp{Khodachenko2012}).

The size of the aurora is calculated according to Eq.~\ref{e_aurora} and the magnetopause stand off distance following the relation \citep{Baumjohann1996}
\begin{equation}
    R_s = \left( \frac{\mu_0 f_0^2 \mathcal{M}_{p}^2}{8 \pi^2 \rho_{\rm sw} v_{\rm sw}^2} \right)^{1/6},
    \label{e_Rs}
\end{equation}
where $\mu_0$ is the diamagnetic permeability of free space, $f_0 \approx 1.22$ is magnetosphere form-factor.

Fig.~\ref{f1} presents the dependence of the emitted soft X-ray power on the orbital distance of the planet. The letters mark the emission levels for HD~209458b estimated above. We should note that we did not take into account the unknown rotation rate of the host star, which leads to an overestimate of the plasma flow speed and, respectively, the emission level in the corotation regime. The letter marks for HD~209458b don't lie exactly on the lines because of the difference between simple estimates used for $R_s$ and $\mathcal{M}_{p}$ to those estimated via comprehensive modelling by \cite{Kislyakova2014}. As a consequence these plots only qualitatively describe the behaviour of the soft X-ray emission due to the SWCX mechanism. For every particular planet, an individual consideration should be made similar to the one above for HD~209458b.

The main conclusion of our results is that the soft X-ray emission is the highest for closest hot Jupiters and strongly depends on the interaction area size (see the difference between the aurora and the whole hemisphere case -- lower and upper red and blue lines, respectively). The simple Equation \ref{e_aurora} \citep{Vidotto2011} can be used only for hot Jupiters and yields a significant overestimate for Jupiter (see the two green lines). For non tidally locked gas giants on wide orbits, the lower green line presents the most plausible estimate.

We should also note that the corotation regime probably breaks closer to the star than shown on Fig.~\ref{f1}. However, this is not so easy to restrict because of many unknown parameters.

Soft X-ray emission from exoplanets orbiting a younger star with a denser stellar wind is always stronger than the emission from an planet embedded in the current solar wind (eq.~\ref{e_int}), which is confirmed by Fig.~\ref{f1}b, simply because the number of emitted photons is proportional to the wind mass flux assuming the same $f$.

%The letters A, B, C, D mark the estimated values for HD~209458b: stellar wind and magnetosphere parameters from \cite{Kislyakova2014}, the interaction in the whole hemisphere with the radius $R_s$ (A); the same magnetosphere parameters, but in the corotational regime (B); interaction only in the auroral region, non-corotational regime (C); interaction only in the aurora, corotational regime (D).

\section{Conclusions}

In this work, we presented a simple estimate of the possible X-ray emission from the close-in gas giants emitted due to the SWCX mechanism. We have shown for the example of HD~209458b that this mechanism alone can be responsible for the X-ray emission intensity of the order of $\approx 10^{22}$~erg~s$^{-1}$, which is $\approx 10^6$ times higher than the X-ray emission from Jupiter.

We have discussed the possibility to observe the soft X-ray flux from close-in extrasolar giant planets and have shown that although this emission exceeds the intensity of the Jovian soft X-ray emission by several orders of magnitude, it is unobservable with present-day facilities because of the large distances to the systems.

%We also show, that this method to some extent can be used to probe the stellar wind properties and the mass loss rate of the host star. Since the intensity of the X-ray emission is directly connected with the amount of the heavy charged ions in the stellar wind, one can estimate the mass loss rate of the star by knowing the X-ray emission and the fraction of the heavy ions in the stellar wind $f$. \textcolor{red}{[Colin - please check this]}.

The main conclusion of the study is that hot Jupiters should be bright X-ray sources in comparison to the Solar system giant planets. The spectrum of this emission as well as the influence of other X-ray producing mechanisms should be the subject of future study.

%\textcolor{red}{\textit{some selfcriticism} Well done!}

%ACKNOWLEDGEMENTS
\acknowledgements{This study was carried out with the support by the FWF NFN project S116601-N16 ''Pathways to Habitability: From Disk  to Active Stars, Planets and Life'' and the related subprojects S116 604-N16 and S116 607-N16. L.F. acknowledges financial support from the Alexander vun Humboldt foundation. L.F. thanks Lorenzo Lovisari for useful discussions. V.Z. acknowledges support from MES RF project 14.Z50.31.0007 "Lab.Astrophysics".}

\end{document}